# Deconfining Chiral Transition in QCD on the Lattice*

K. Kanaya†
*Institute of Physics, University of Tsukuba, Ibaraki 305, Japan*

**Abstract**

The deconfining chiral transition in finite-temperature QCD is studied on the lattice using Wilson quarks. After discussing the nature of chiral limit with Wilson quarks, we first study the case of two degenerate quarks $N_F = 2$, and find that the transition is smooth in the chiral limit on both $N_t = 4$ and 6 lattices. For $N_F = 3$, on the other hand, clear two state signals are observed for $m_q \lesssim 140$ MeV on $N_t = 4$ lattices. For a more realistic case of $N_F = 2+1$, i.e. two degenerate u and d-quarks and a heavier s-quark, we study the cases $m_s \simeq 150$ and 400 MeV with $m_u = m_d \simeq 0$: In contrast to a previous result with staggered quarks, clear two state signals are observed for both cases, suggesting a first order QCD phase transition in the real world.

*Talk presented at International Workshop *"From Hadronic Matter to Quark Matter: Evolving View of Hadronic Matter"*, Oct. 30 – Nov. 1, 1994, Yukawa Institute for Theoretical Physics, Kyoto, Japan. To appear in the proceedings.

†in collaboration with Y. Iwasaki, S. Kaya, S. Sakai, and T. Yoshié.

# 1 INTRODUCTION

The study of QCD on the lattice has shown that QCD has a finite temperature phase transition/crossover at $T \sim 100 - 200$ MeV. In the light quark limit (chiral limit), the chiral symmetry, that is broken spontaneously at low temperatures, is resored when $T$ gets sufficiently high. In the heavy quark limit (pure gauge limit), existence of a deconfining phase transition separating the low temperature confining phase and the high temperature deconfining phase is established. Studies on the lattice for intermediate quark masses indicate that the chiral transition in the light quark limit and the deconfining transition in the heavy quark limit are two ends of a single transition line with a probable crossover part in the middle. This transition/crossover plays a primary role in the studies of the quark-gluon-plasma state in heavy ion collisions and in the evolution of early Universe. A decisive information for these studies is the order of the transition. Therefore it is important to pin down the nature of the deconfining chiral transition with the values of parameters corresponding to the real world.

Previous investigations based on universality arguments as well as numerical studies on the lattice show that the order of the transition in the chiral limit depends on the number of flavors $N_F$: With degenerate $N_F$ quarks, the chiral transition is first order for $N_F \geq 3$, while a continuous transition is suggested for $N_F = 2$. Because the s-quark mass $m_s$ is of the same order of magnitude as the transition temperature, the order of the transition may depend on the value of $m_s$ sensitively. We need to include the s-quark to make a prediction for the real world.

In this report, we study the finite temperature transition on the lattice using Wilson fermions [1] as lattice quarks. The action is given by

$$S = S_g + S_W \tag{1}$$

$$S_g = -\frac{\beta}{6} \sum_{\text{plaquettes}} \text{Tr}(P_p + P_p^\dagger) \tag{2}$$

$$S_W = \sum_{f=1}^{N_F} \sum_{x,y} \bar{\Psi}_x^f D_{x,y} \Psi_y^f \tag{3}$$

where $\beta = 6/g^2$ with gauge coupling $g$,

$$D_{x,y} = \delta_{x,y} - K_f \sum_{\mu=1}^{4} \{(1-\gamma_\mu) U_{x,\mu} \delta x + \mu, y + (1+\gamma_\mu) U_{y,\mu} \delta x, y + \mu\},$$

and $K_f$ is the hopping parameter to tune the quark mass for the $f$-th flavor. With SU(3) gauge group lattice spacing $a$ is a decreasing function of $\beta$ and the temperature is given by $T = 1/(N_t a(\beta))$ with $N_t$ the lattice size in the temporal direction.

In the next section, we first discuss the chiral limit with Wilson quarks. We then present the results of our simulations for the case of degenerate quarks with $N_F = 2$ and 3. Finally we discuss a more realistic case of $N_F = 2+1$: two degenerate light u and d-quarks and a heavier s-quark. Comparison with previous results using staggered quarks is also made. We mainly perform simulations on $8^2 \times 10 \times N_t$ and $12^3 \times N_t$ lattices with $N_t = 4$. For $N_F = 2$ we also study the case of $N_t = 6$. Preliminary reports of the results presented here can be found in refs. [2, 3, 4, 5] and [6]. More detailed reports will be available soon.

# 2 Chiral limit

Before studying the finite temperature transition, let us discuss the definition of chiral limit with Wilson quarks. In the continuum limit $\beta = \infty$, it is easy to see that the quark masses vanish at $K = 1/8 = K_C(\beta = \infty)$ as $m_q \propto 1/K - 1/K_C$. When the system is in the chirally broken phase, we expect $m_\pi^2 \propto m_q$ and $K_C$ can also be defined as the point where $m_\pi^2$ vanishes. Let us denote $K_C$ defined through $m_q$ as $K_C^{m_q}$ and that through $m_\pi^2$ as $K_C^{m_\pi^2}$ where a distinction among these two definitions is needed. For finite lattice spacings $a > 0$, however, Wilson's action (3) breaks the chiral symmetry explicitly through the Wilson term and the notion of chirality as well as the definition of $K_C$ becomes unclear.

Nevertheless numerical simulations at finite $\beta$'s show that, in the low temperature confining phase (with $N_F \leq 6$), $m_\pi^2$ vanishes linearly in $1/K$ at finite $K_C(\beta)$.[1] Accordingly, a careful perturbative study of Wilson quarks at large $\beta$'s show that, when renormalization counter terms are chosen properly,

[1] In the space of coupling parameters, it is difficult to simulate the points where $m_\pi$ is

axial Ward identities hold also on the lattice to the price of $O(a)$ correction terms [7]. We can therefore consider, for example, a relation:

$$2m_q < 0\,|\,P\,|\,\pi >= -m_\pi < 0\,|\,A_4\,|\,\pi > +O(a) \tag{4}$$

to define $m_q$ where $P$ is the pseudoscalar density and $A_4$ the fourth component of the local axial vector current. Here, the chiral limit $K_C$ can be defined where $m_q$ vanishes. $K_C$ gets a $\beta$-dependence through counter terms.

Independently Tsukuba group [8] proposed to define quark mass through an axial Ward identity neglecting $O(a)$ terms in (4). Maiani and Martinelli [9] also studied $m_q$ defined similarly using axial Ward identity. Numerical studies of this $m_q$ for intermediate $\beta$ ($\beta = 5.85$ with quenched QCD [10] and $\beta = 5.5$ with $N_F = 2$ full QCD [11]) show that (i) $m_q$ vanishes linearly in $1/K$ at finite $K = K_C^{m_q}$, (ii) the value of $m_q$ (therefore the value of $K_C^{m_q}$ also) is almost independent of $T$ and the phase, and (iii) $K_C^{m_q}$ is consistent with $K_C^{m_\pi^2}$ determined with $m_\pi^2$ in the confining phase (Fig. 1). Therefore we have

$$K_C^{m_q}(\mathrm{high}\,T) \simeq K_C^{m_q}(\mathrm{low}\,T) \simeq K_C^{m_\pi^2}(\mathrm{low}\,T).$$

These nice features of our $m_q$ suggest that $O(a)$ chiral violation are small in (4) down to these values of $\beta$. We also note that we can simulate the system in the deconfining phase even just on the $K_C$-line without encountering the critical slowing down. Our $m_q$ therefore provides us a numerically easy way to determine $K_C$ for $\beta \gtrsim 5.5$.

Equivalence of $K_C^{m_q}$ and $K_C^{m_\pi^2}$ can be shown also in the strong coupling limit $\beta = 0$ [12]: In his pioneering paper [1], K. Wilson carried out a strong coupling calculation of meson masses for $T = 0$ without quark loops. We can extend this calculation to our $m_q$ and we get

$$\cosh(m_\pi a) \;=\; 1 + \frac{(1-16K^2)(1-4K^2)}{4K^2(2-12K^2)} \tag{5}$$

$$2\,m_q a \;=\; m_\pi a \frac{4K^2 \sinh(m_\pi a)}{1-4K^2\cosh(m_\pi a)} \tag{6}$$

Our numerical data for $N_F = 2$ at $\beta = 0$ agrees with these formulas well as shown in Fig. 2. (Results for $N_F = 18$ is presented in ref. [12] where the very small due to the critical slowing down. Here, $K_C^{m_\pi^2}$ is defined by a linear extrapolation of $m_\pi^2$ obtained in the confining phase on a lattice as large as possible.

agreement in this case is shown. The rho meson mass, the nucleon mass, and the delta mass also agree with corresponding mass formulas.) We note that both $m_\pi$ and $m_q$ vanish at the same $K = 0.25$ and $m_\pi^2 \propto m_q$ near this $K_C$.

For $0 < \beta \lesssim 5.3$, however, an additional complexity comes on the stage: $m_q$ shows a strange dip when we increase $K$ over the deconfining chiral transition/crossover point $K_T$ and $m_q$ in the deconfining phase is no more independent of $T$ for these $\beta$'s [2, 12, 13, 14]. This strange behavior can be attributed to an effect of $O(a)$ chiral violation of Wilson fermions. In the confining phase, on the other hand, no such strange behavior is found. This interpretation including the smallness of O($a$) effects in the confining phase is confirmed by our recent study with an RG improved action [4, 5]. For these $\beta$'s, we define $K_C$ by a linear extrapolation of $m_\pi^2$ or $m_q$ obtained in the confining phase on a lattice as large as possible.

Fig.3 summarizes the numerical results of $K_C$ for $N_F = 2$ including the data from other collaborations [13, 14, 15, 16]. The sharp change of $K_C$ around $\beta \approx 5.0$ signals the crossover from the strong coupling region to the weak coupling region. Our data for $\beta \leq 4.3$ are obtained mainly on an $N_t = 4$ lattice. Small difference in the value of $K_C^{m_\pi^2}$ and $K_C^{m_q}$ there (at most of the order of 0.01) will be a result of errors in the extrapolation to $m_\pi^2 = 0$ and $2m_q = 0$, and also of the O($a$) chiral violation by Wilson fermions — our data is not accurate enough yet to decide the main cause among these possibilities. Taking the errors due to the difference of $K_C^{m_\pi^2}$ and $K_C^{m_q}$ into account, we find that $N_t$ (as well as $N_F$) dependence of $K_C$ is small for $\beta \lesssim 4.5$.

In summary, in spite of the explicit chiral violation, Wilson quarks show rather clean chiral behaviors on the lattice. Although some care is required in handling numerical results of $m_q$ etc. in the deconfining phase at small $\beta$'s, most effects of chiral violation seem to be absorbed by renormalizations such as a shift of $K_C$ and $Z$-factors for axial currents etc.

# 3 $N_F = 2$

The location of $K_T$ can be easily identified numerically by rapid changes of physical observables such as plaquette, Polyakov loop, hadron masses, and $m_q$. Close to the chiral limit $K_C$, however, the critical slowing down in the confining phase makes the simulation practically impossible. In numerical simulations of QCD, this critical slowing down appears as an rapid increase of the number $N_{\rm inv}$ of CG iterations to invert the quark matrix, closely connected with the appearance of very small eigen values of quark matrix [11]. As discussed before, on the other hand, $N_{\rm inv}$ remains small in the deconfining phase even just on the $K_C$-line. We can therefore study the chiral transition by simulations along the $K_C$-line in the deconfining phase: Decreasing $\beta$ from the deconfining phase, the location of the chiral transition, i.e. the crossing point $\beta_{CT}$ of $K_T$ and $K_C$ lines, can be determined by a sudden increase of $N_{\rm inv}$. The value of $\beta_{CT}$ determined by this method is consistent with an extrapolation of the $K_T$-line determined by the conventional method as shown in the phase diagrams given below. The nature of the transition there can also be studied by monitoring physical observables on the $K_C$-line.

In Refs. [3, 6, 17] this method is applied to the case of $N_F = 2$. Phase diagram for $N_F = 2$ is shown in Fig. 4. Time history of $N_{\rm inv}$ on the $K_C$-line is given in Fig. 5 for $N_t = 4$. Time histories of plaquette, Polyakov loop, $m_\pi$ etc. at the same points confirm that the system is developing into a confined state for $\beta \leq 3.9$. Therefore we conclude that $\beta_{CT} \simeq 3.9 - 4.0$ for $N_t = 4$. Similar study on an $N_t = 6$ lattice gives $\beta_{CT} \simeq 4.0 - 4.2$ and a study on a $18^3 \times 24$ lattice gives $\beta_{CT} \sim 4.5 - 5.0$.

Our result of $m_\pi^2$ on the $K_C$-line is shown in Fig. 6. For both $N_t = 4$ and 6, we find that, when we decrease $\beta$ toward $\beta_{CT}$, $m_\pi^2$ decreases rapidly and is consistent with zero at $\beta_{CT}$. This suggests that the transition is smooth in the chiral limit, in accord with the result of an universality argument [18] and finite-size scaling studies with staggered quarks [19, 20] suggesting a second order transition. Off the chiral limit, $m_q > 0$, we may therefore expect that $K_T$ is crossover. [2]

[2]Concerning the nature of the transition on the $K_T$-line at intermediate values of $K$, MILC collaboration reported recently an unexpected phenomenon that the transition/crossover becomes once very strong when we increase $m_q$, while it becomes weak again when we further increase $m_q$ [13, 14]. For $N_t = 4$, $K_T$ is very strong at $K \sim 0.18$. For $N_t = 6$ they even found first order signals at several $K$'s. Consulting the phase diagram Fig. 4, we note that these values of $K$ just correspond to the region where the $K_C$-line shows a sharp bend due to the crossover phenomenon between weak and strong coupling regions. The $K_T$-line that deviates from the $K_C$-line at $\beta_{CT}$ gets very close to the $K_C$-line again there. It is therefore plausible that the strong $K_T$ is a lattice artifact in the crossover region for these $N_t$ caused by this unusual relation of $K_C$ and $K_T$ lines. Close to the continuum limit with a large enough $N_t$, the $K_C$-line is smooth and the $K_T$-line simply gets apart from it when we increase $m_q$.

# 4 $N_F = 3$

We apply the same method to study the chiral transition for $N_F = 3$ and we find a two-state signal at $\beta_{CT} \simeq 3.0$ at $N_t = 4$ [3, 17]: When we start from a deconfining configuration prepared with a larger $\beta$, it remains stable in the deconfining phase with small $N_{\rm inv}$. However, when we start from a mixed initial configuration, $N_{\rm inv}$ increases rapidly with MC-time and simultaneously plaquette, Polyakov loop, and $m_\pi^2$ decrease rapidly, showing that the system is developing into a confining state. This implies a first order chiral transition for $N_F = 3$. Similar first-order signal is obtained also for $N_F = 6$ [2]. These results are consistent with the prediction based on universality [18]. Our results of $m_\pi^2$ on the $K_C$-line is summarized in Fig. 7.

We now extend the study to finite $m_q$'s by increasing $\beta$ on the $K_T$-line. Our phase diagram for $N_F = 3$ is given in Fig. 8. We find clear two-state signals for $\beta = 4.0$, 4.5 and 4.7 on both $8^2 \times 10 \times 4$ and $12^3 \times 4$ lattices (Fig. 9). However, when we further increase $\beta \geq 5.0$, we find no clear signs of metastability. The critical value of the quark mass $m_q^{\rm crit}$, up to which we get a clear first order signal on these lattices, is bounded from below by $m_q$ for $\beta = 4.7$. We find $m_q^{\rm crit} a \geq 0.175(2)$ and $(m_\pi/m_\rho)^{\rm crit} \geq 0.873(6)$.

In order to convert the critical quark mass into physical units, we use the results of $m_\rho a$ and $m_q a$ for these $\beta$'s. When we plot the masses in the confining phase as a function of $1/K - 1/K_C$, they are almost independent of $\beta$, $N_F$ and $N_t$ for $\beta \lesssim 4.7$. Fig. 10 shows our results for $N_F = 2$ and 3. Identifying $m_\rho(K_C)$ with 770 MeV we get $a \simeq 0.8$ GeV$^{-1}$ for $\beta \lesssim 4.7$. It should be noted that the vector meson mass "$m_\rho$" for $m_q \simeq 150$ MeV is consistent with $m_\phi = 1020$ MeV ($\phi \approx s\bar{s}$ due to the approximate "ideal mixing"), indicating that our definition of $m_q$ is close to that of the current quark mass.

From this value of $a$ we get $m_q^{\rm crit} \gtrsim 140$ MeV. We note that our $m_q^{\rm crit}$ is much larger than that obtained previously with staggered quarks at $N_t = 4$ with spatial $8^3$ – $16^3$ lattices [21, 22]: two state signals are observed for $m_q a = 0.025$, while no clear metastabilities are found for $m_q a = 0.075$. Using a result for meson masses at corresponding $\beta$ [23] (the values for $N_F = 4$, because they are only available) we obtain $m_q^{\rm crit} \simeq 12$ – 38 MeV and $(m_\pi/m_\rho)^{\rm crit} \simeq 0.42$ – 0.58 for staggered quarks.

## 5 $N_F$ = 2+1

We now study the case of $N_F = 2+1$: $m_u = m_d < m_s$. If $m_s \gg T_c$ then s-quark will decouple from the dynamics relevant to the deconfining transition and we expect a smooth deconfining transition similar to the case of $N_F = 2$. On the other hand, when $m_s \ll T_c$, the results for $N_F = 3$ implies that the transition is first-order. Realistic value for $m_s$ seems to be of the same order as $T_c$. Therefore we should include the dynamics of s-quark properly. In this connection, we note that, when $m_s < m_q^{\rm crit}$ (the critical quark mass for the case of degenerate $N_F = 3$), we can expect a first-order deconfining transition, while when $m_{u,d} > m_q^{\rm crit}$, we should expect a crossover. Recall the results for $m_q^{\rm crit}$ discussed in the previous section: $m_q^{\rm crit} \gtrsim 140$ MeV with Wilson quarks and $m_q^{\rm crit} \simeq 12 - 38$ MeV with staggered quarks. Our large $m_q^{\rm crit}$ opens the possibility of a first order deconfining transition at quark masses corresponding to the real world.

In order to see what happens with realistic quark masses, we study the cases $m_s \simeq 150$ MeV and 400 MeV. Corresponding values of $K_s$ are obtained from Fig. 10. For u and d-quarks we fix $m_u = m_d \simeq 0$. We apply the method which we used in previous sections to study the chiral transition: Keeping $K_u = K_d = K_C$ and $K_s$ at the point corresponding to the case of $m_s \simeq 150$ or 400 MeV, we decrease $\beta$ until we hit $K_T$ and study the nature of the transition there. Fig. 11 is an example of time history we get. We find two state signals on an $8^2 \times 10 \times 4$ lattice for both $m_s = 150$ and 400 MeV and also on a $12^3 \times 4$ lattice for $m_s = 400$ MeV. Our results of $m_\pi^2$ on the $K_C$-line are shown in Fig. 7.

With staggered quarks, Columbia group [22] reported that no transition occurs at $m_u a = m_d a = 0.025$ and $m_s a = 0.1$ ($m_{u,d} \simeq 12$ MeV, $m_s \simeq 50$ MeV using the value of $a$ from [23]). Although a direct comparison with our result is difficult because of the difference in simulation parameters, this implies a discrepancy among lattice quark formulations for these $\beta$'s and/or a sensitive dependence of the nature of $K_T$ on the value of light quark masses $m_u$ and $m_d$.

# 6 CONCLUSION

We have studied the nature of the finite temperature transition with Wilson quarks for $N_F = 2$, 3 and 2+1. For $N_F = 2$ the chiral transition is smooth on an $N_t = 6$ lattice in accord with our previous result at $N_t = 4$. For $N_F = 3$ at $N_t = 4$, clear two state signals are observed on the $K_T$-line for $m_q \lesssim 140$ MeV. For $N_F = 2 + 1$ we have studied the cases $m_s \simeq 150$ and 400 MeV with $m_u = m_d \simeq 0$, and we have found two state signals for both cases. This suggests a first order finite temperature transition in the real world.

I am grateful to my collaborators Y. Iwasaki, S. Kaya, S. Sakai, and T. Yoshié, with whom the studies reported here are done. Simulations are performed with HITAC S820/80 at KEK and with QCDPAX and VPP-500/30 at the University of Tsukuba. I thank members of KEK and the other members of QCDPAX collaboration for their support. This work is in part supported by the Grant-in-Aid of Ministry of Education, Science and Culture (No.06NP0601).

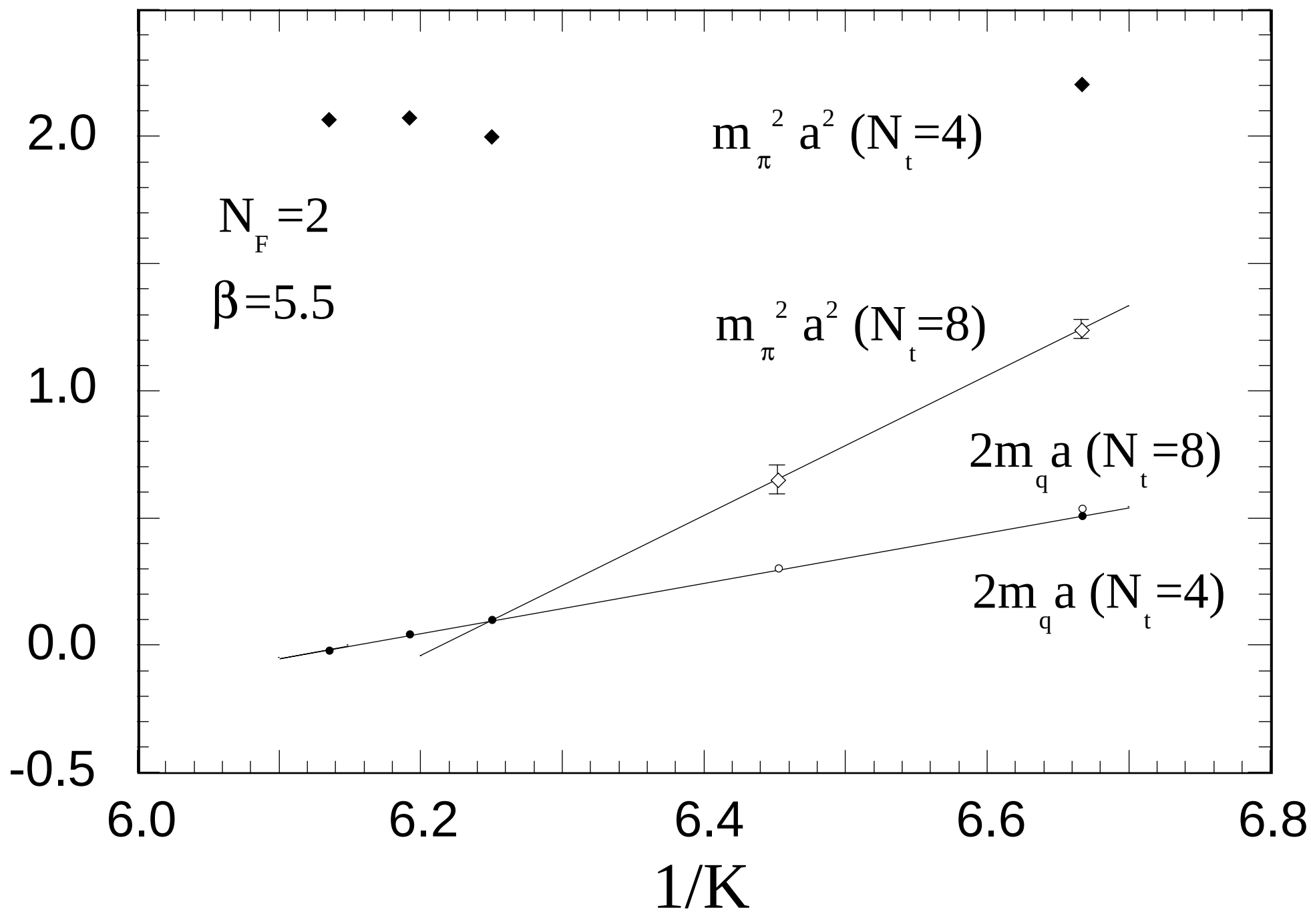

Figure 1: $m_\pi^2 a^2$ and $2m_q a$ for $N_F = 2$ and $\beta = 5.5$ on an $8^2 \times 20 \times N_t$ lattice with $N_t = 4$ (filled symbols) and 8 (open symbols) [11]. Straight lines are the results of a linear fit for $2m_q(N_t = 4)$ and $m_\pi^2(N_t = 8)$. For this $\beta$ and these values of $K$, $N_t = 4$ means that the system is in the high temperature deconfining phase, while $N_t = 8$ means the low temperature confining phase.

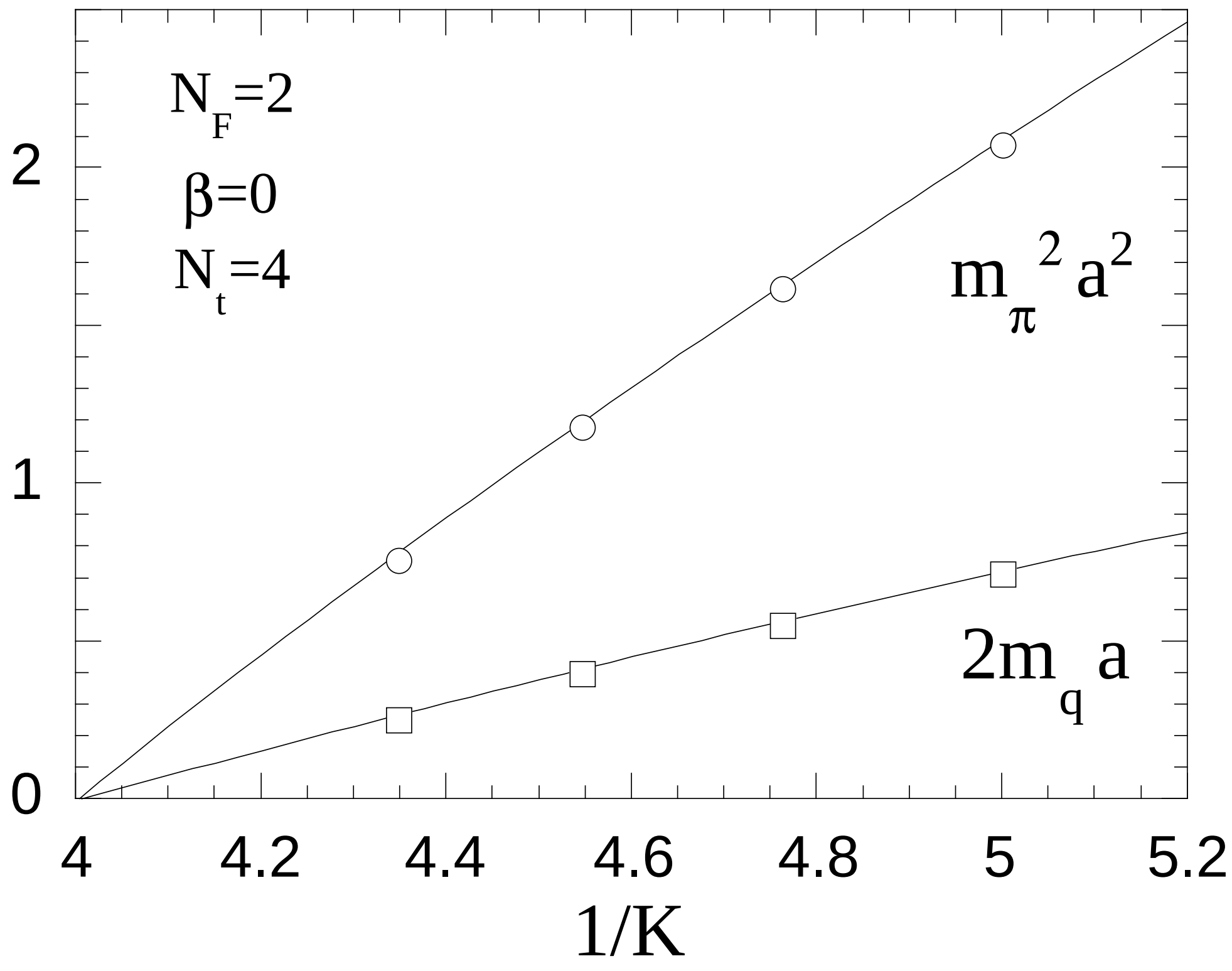

Figure 2: The same as Fig. 1 for $\beta = 0$ on an $8^2 \times 10 \times 4$ lattice. Errors are smaller than the size of symbols. Solid curves are the results of a strong coupling calculation (6).

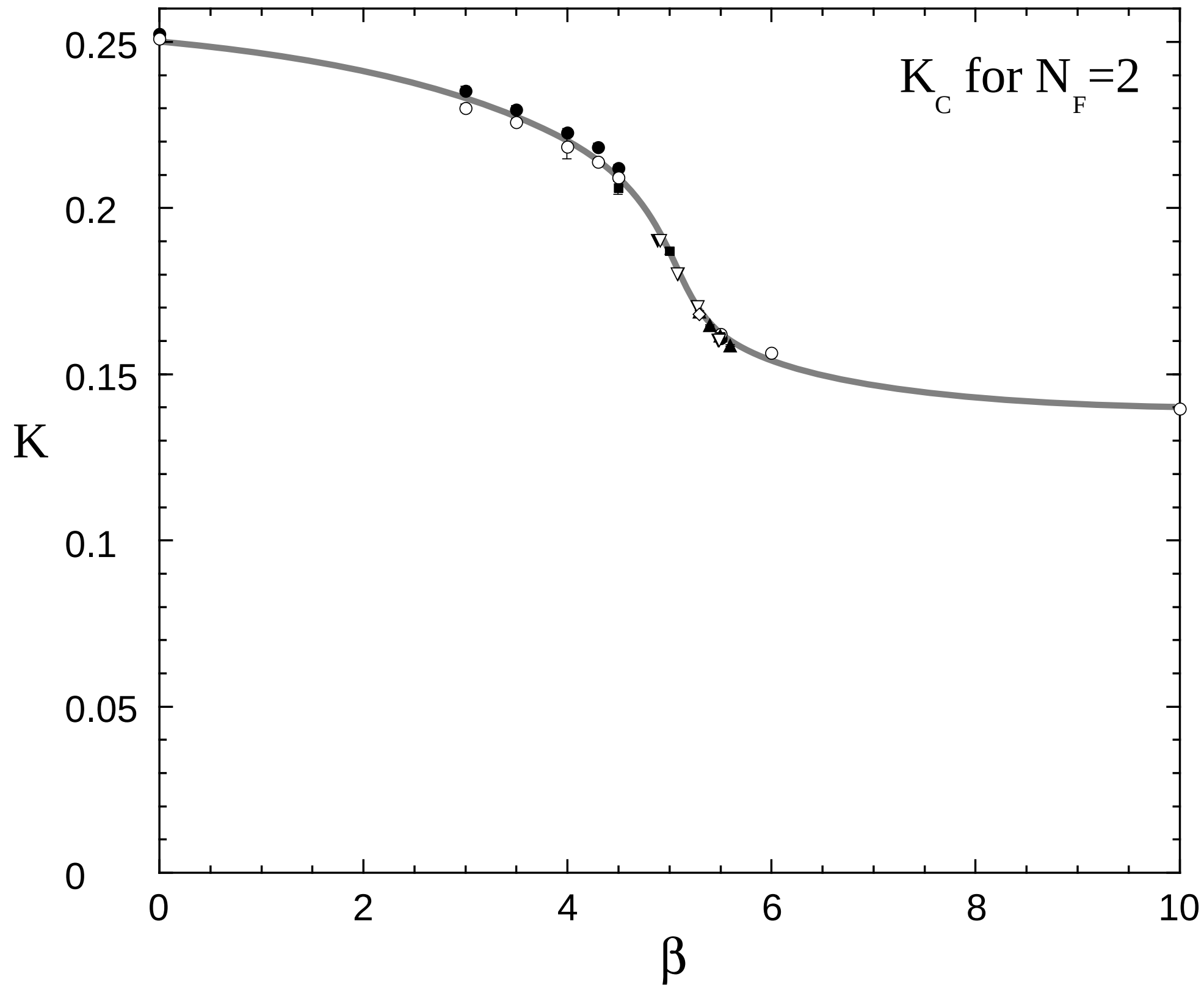

Figure 3: The line of $K_C$ for $N_F = 2$. Filled symbols are for $K_C^{m_\pi^2}$ and open symbols are for $K_C^{m_q}$. Circles are for our data. Our $K_C^{m_q}$ data at $\beta$=6.0 and 10.0 are determined in the deconfining phase.

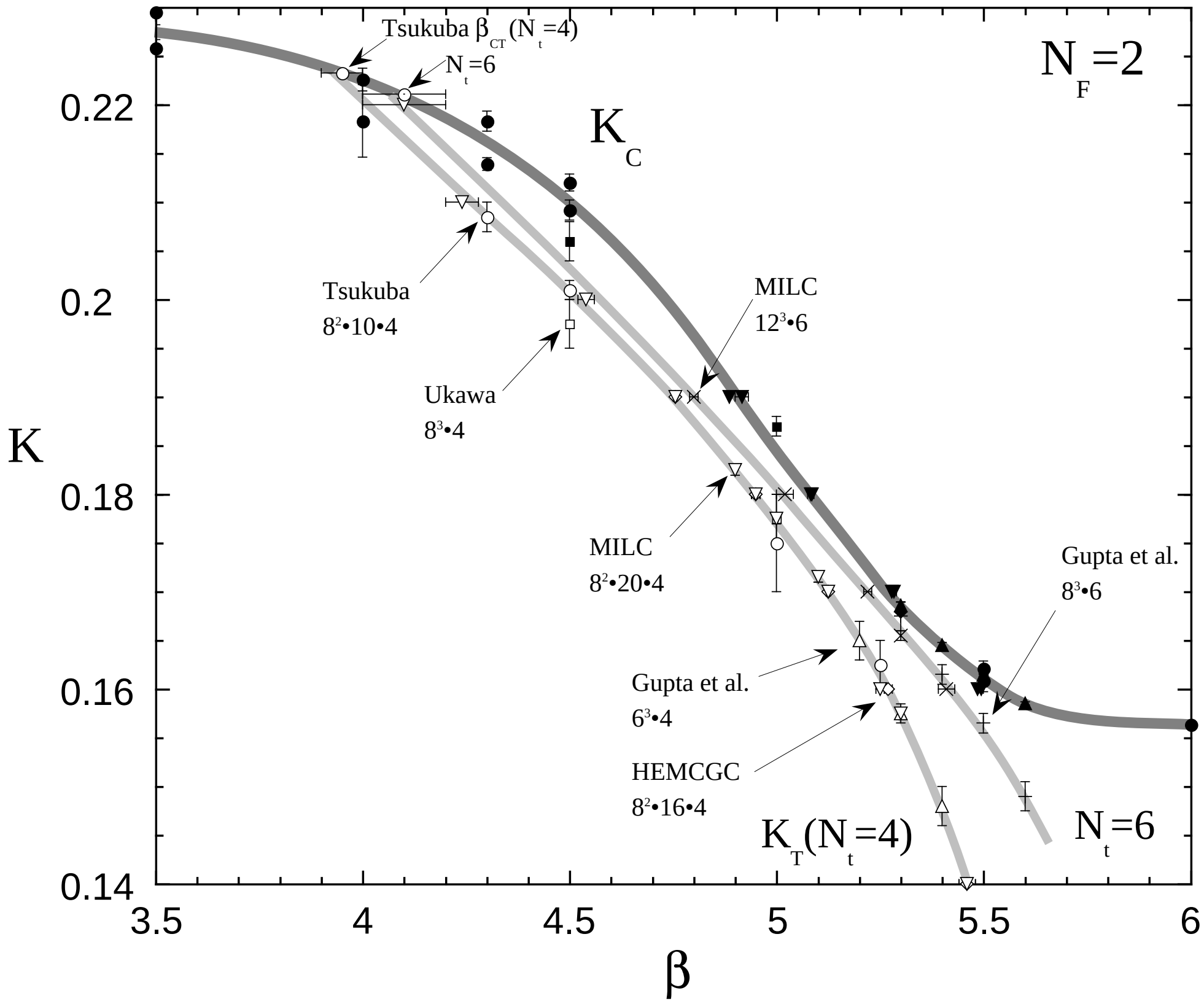

Figure 4: Phase diagram for $N_F = 2$ near $K_C$. $K_T$-lines are for $N_t = 4$ and 6.

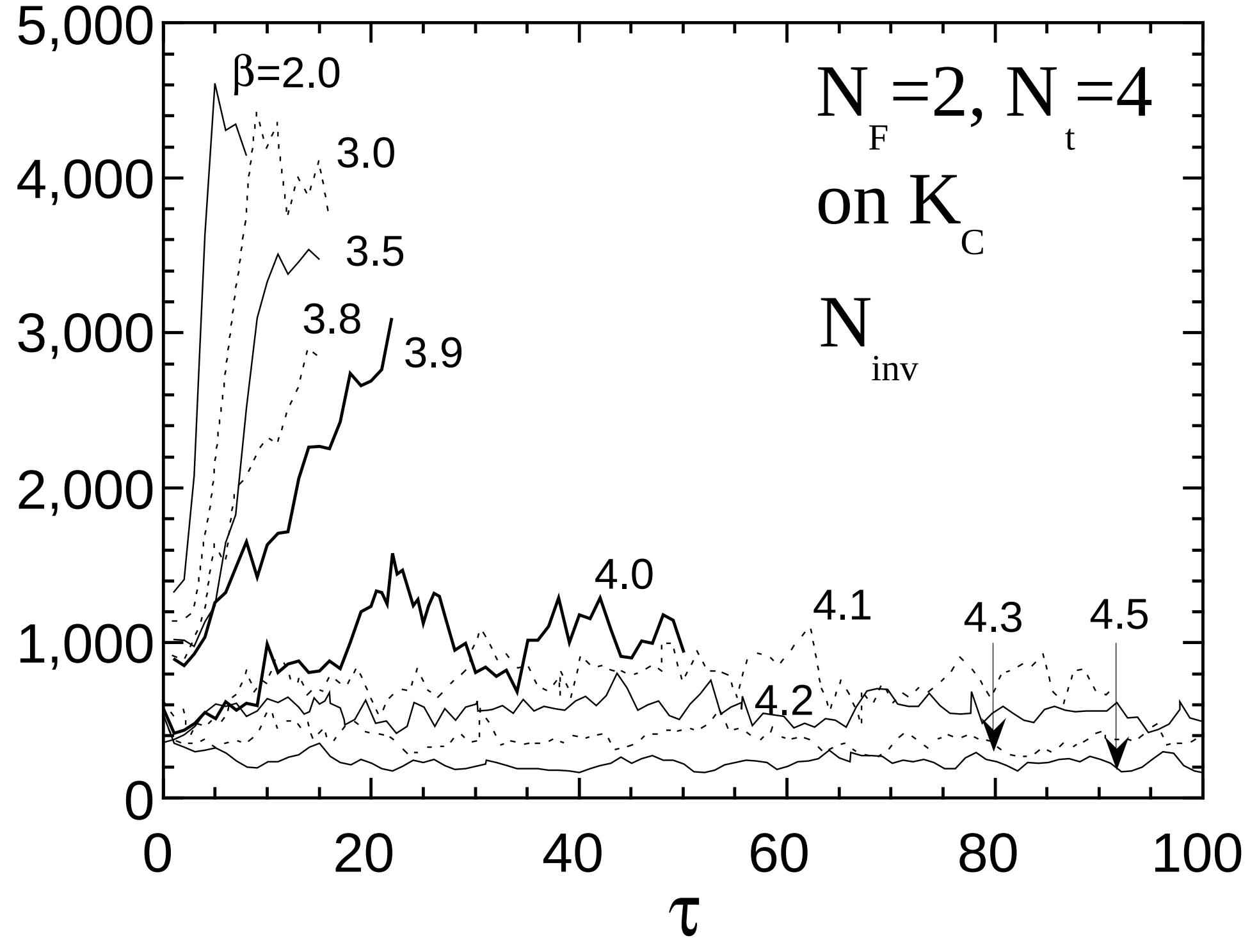

Figure 5: Time history of $N_{\rm inv}$ on the $K_C$-line for $N_F = 2$ on an $8^2 \times 10 \times 4$ lattice.

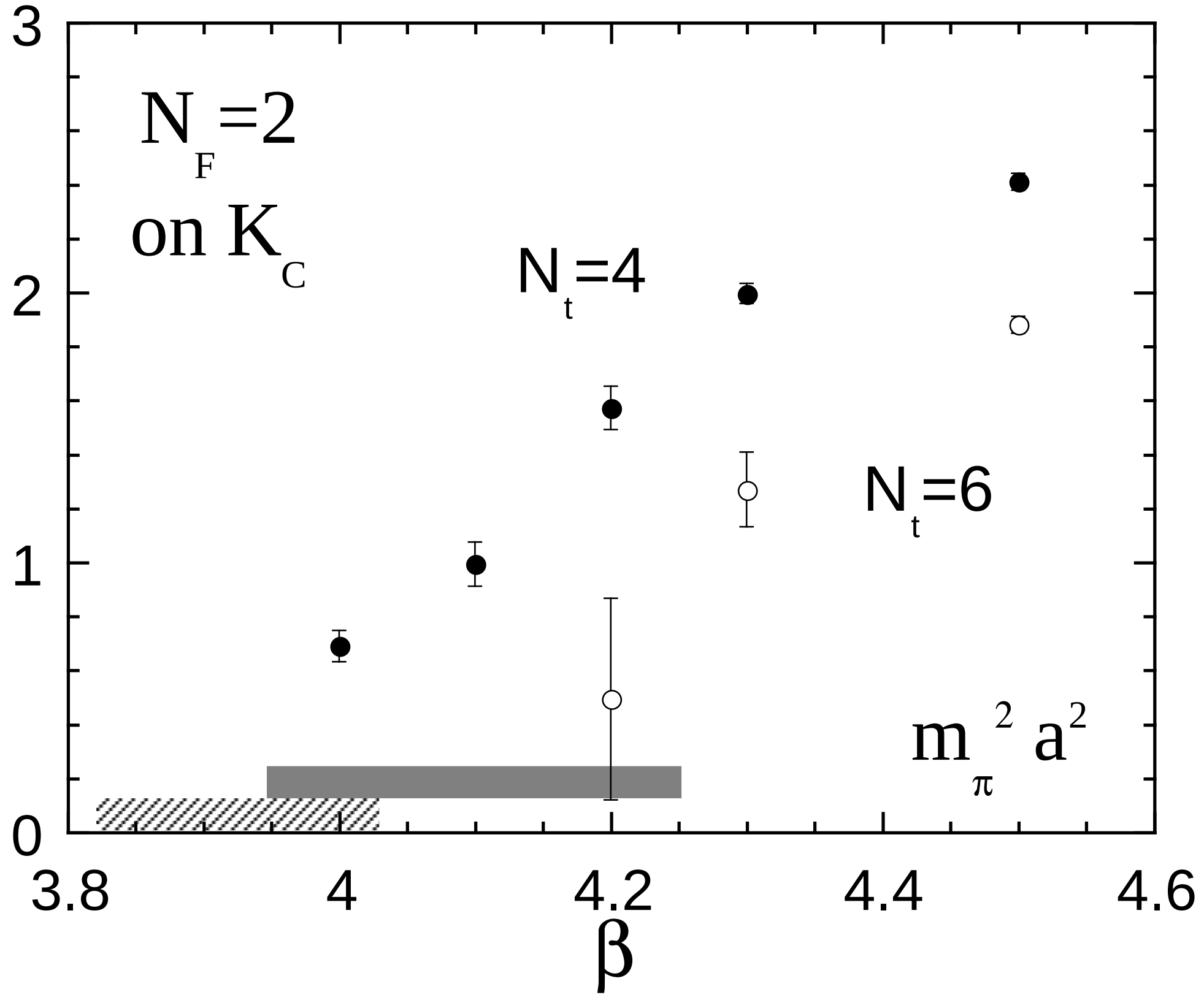

Figure 6: $m_\pi^2 a^2$ on the $K_C$-line for $N_t = 4$ and 6 with spatial $8^2 \times 10$ lattice. The locations of the crossing point $\beta_{CT}(N_t)$ are shown by shaded bars.

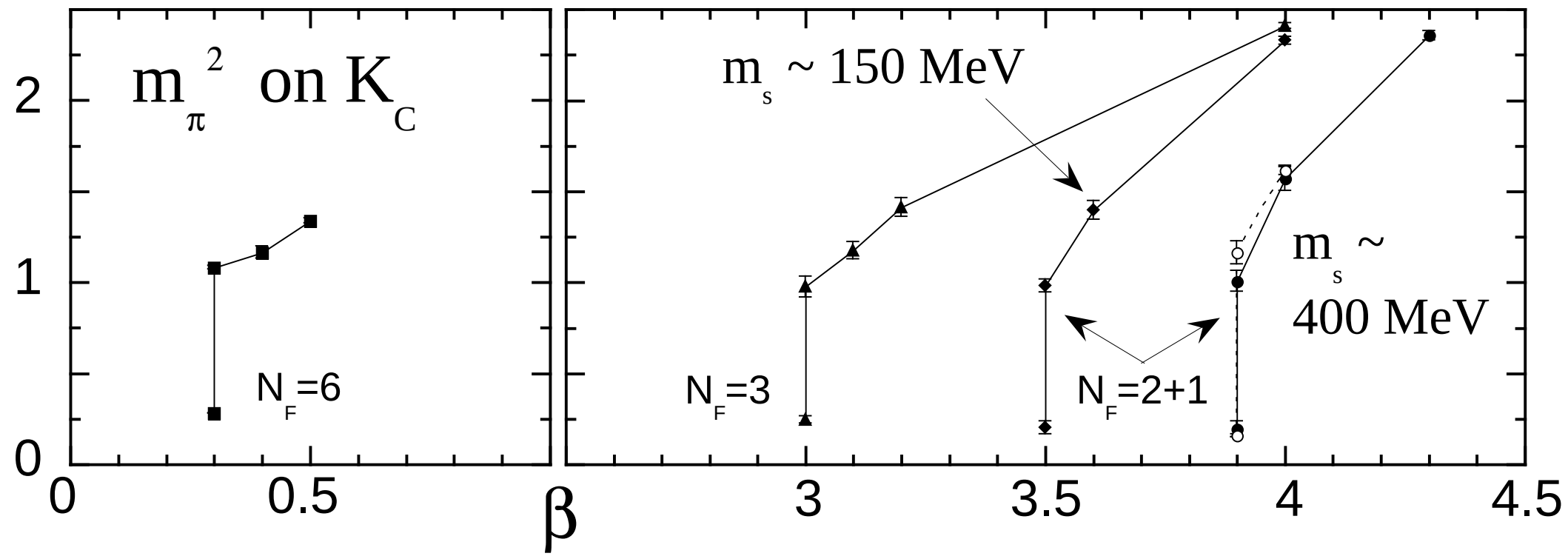

Figure 7: $m_\pi^2 a^2$ on the $K_C$-line for $N_F$ =6, 3 and 2+1. The filled symbols are for a $8^2 \times 10 \times 4$ lattice, while the open ones are for a $12^3$ lattice. The lower points at $\beta_{CT}$ represent the upper bounds for $m_\pi^2$ obtained by time-histories from mixed initial configurations.

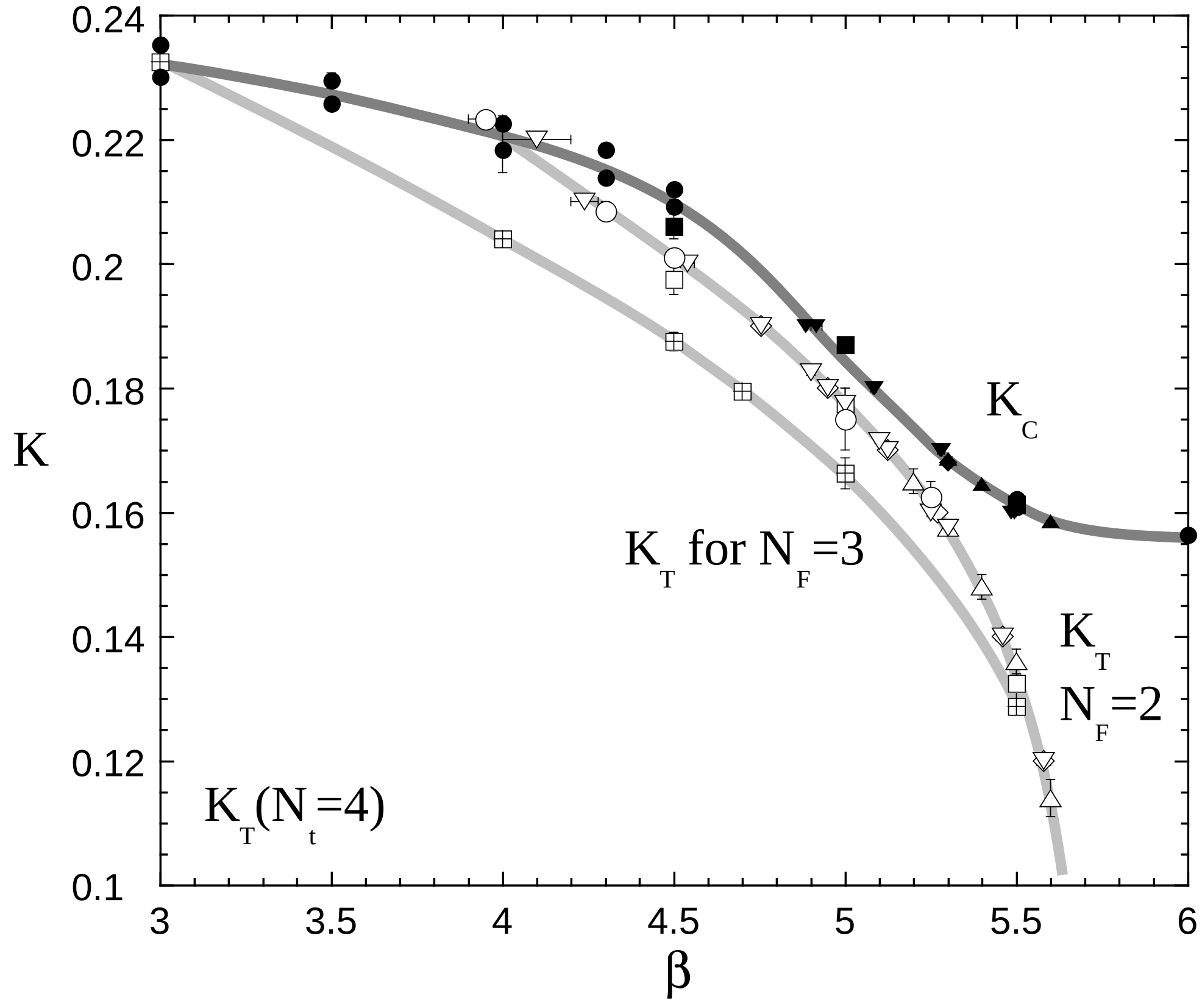

Figure 8: Phase diagram for $N_F = 3$ at $N_t = 4$ near $K_C$. $K_T$-line for $N_F = 2$ is also plotted.

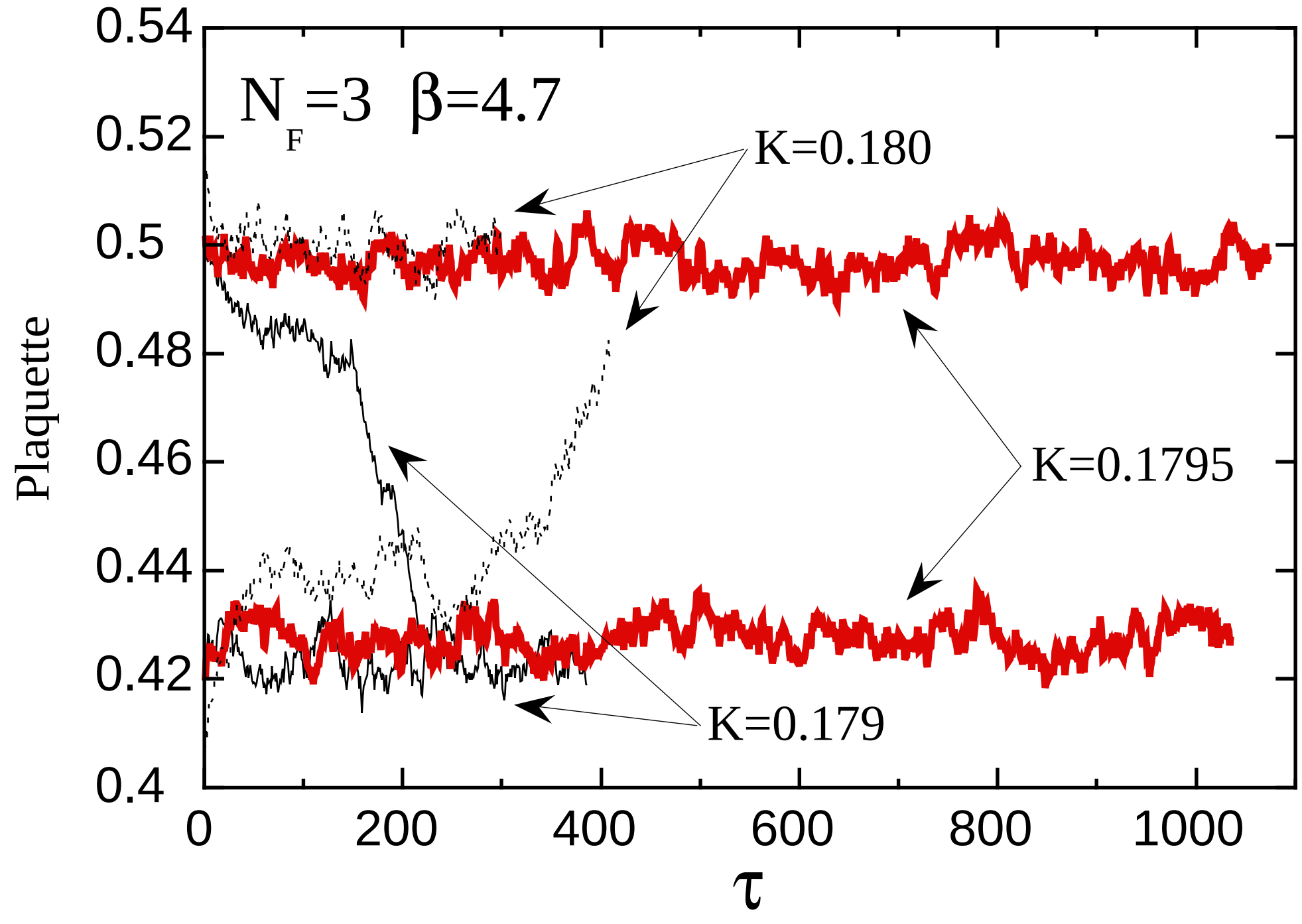

Figure 9: Time histories of the plaquette for $N_F = 3$ at $\beta = 4.7$ on a $12^3 \times 4$ lattice.

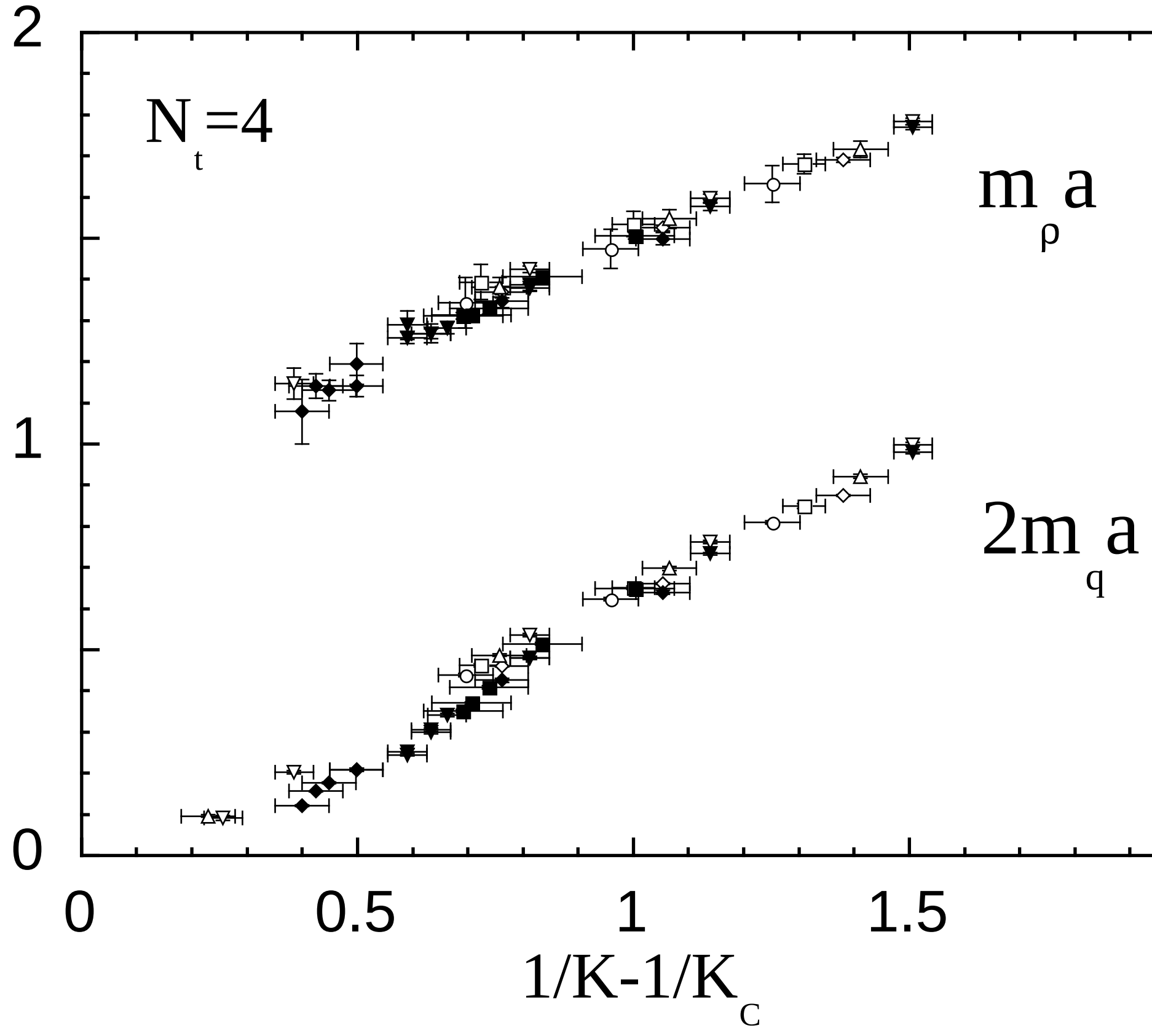

Figure 10: $m_\rho a$ and $2m_q a$ in the confining phase. Open symbols are for $N_F = 2$, $\beta = 3.0$, 3.5, 4.0, 4.3, and 4.5 on an $8^2 \times 10 \times 4$ lattice. Filled symbols are for $N_F = 3$, $\beta = 4.0$, 4.5 and 4.7 on $8^2 \times 10 \times 4$ and $12^3 \times 4$. $K_C(\beta)$ is determined by $m_\pi^2$ and $m_q$ for $N_F = 2$, with errors taking into account the difference between two definitions.

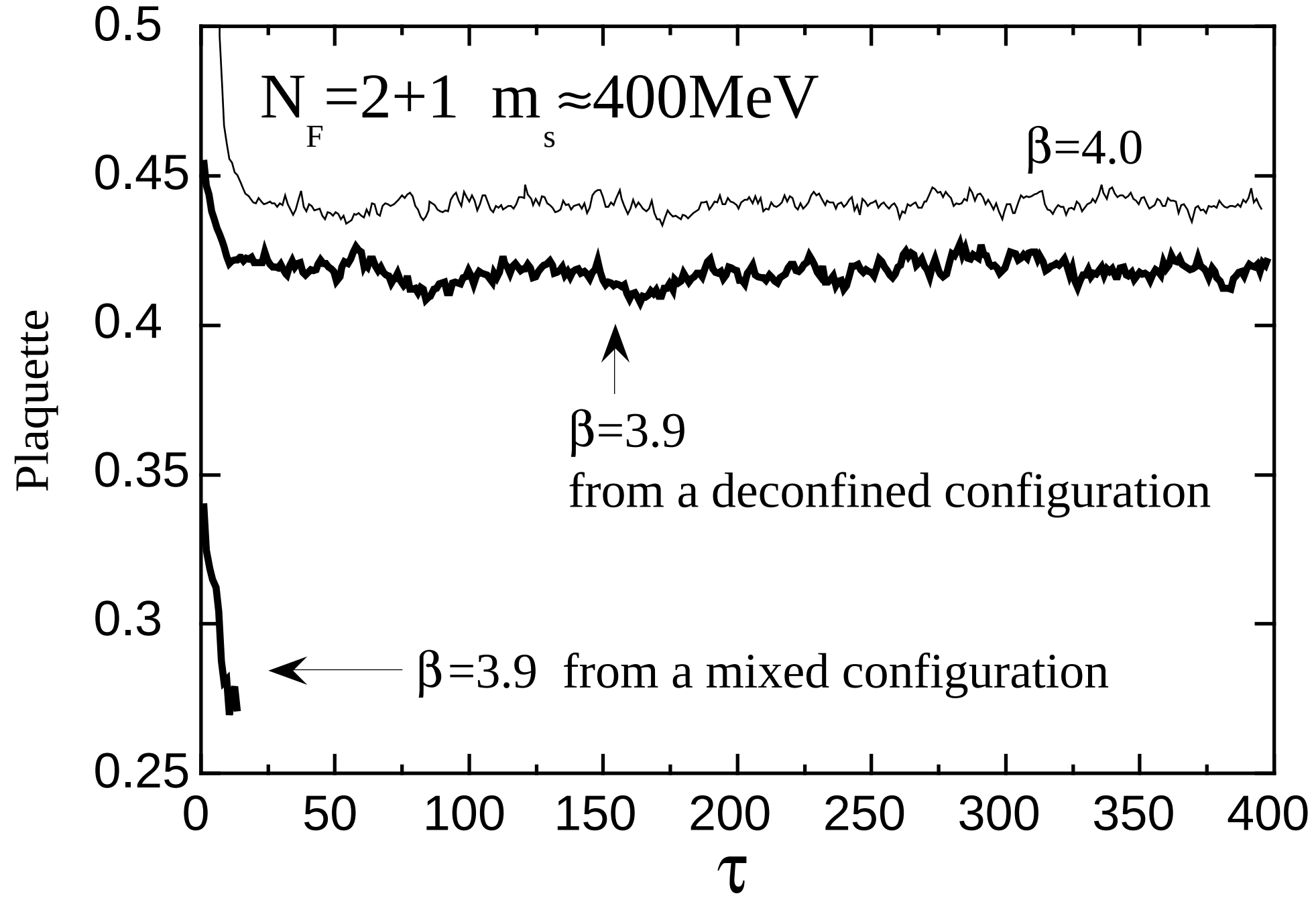

Figure 11: Time histories of the plaquette for $N_F = 2+1$ with $m_s \simeq 400$ MeV on a $12^3 \times 4$ lattice.